\documentclass[final,1p,times]{elsarticle} % do not change
\usepackage{graphicx} % do not change
\usepackage{epsfig}
\usepackage{amssymb} % do not change
\usepackage{amsthm} % do not change
\usepackage{lineno} % do not change

\newcommand{\be}{\begin{equation}}
\newcommand{\ee}{\end{equation}}
\newcommand{\bea}{\begin{eqnarray}}
\newcommand{\eea}{\end{eqnarray}}
\newcommand{\lton}{\mathrel{\lower.9ex \hbox{$\stackrel{\displaystyle 
<}{\sim}$}}}  
\newcommand{\gton}{\mathrel{\lower.9ex \hbox{$\stackrel{\displaystyle 
>}{\sim}$}}}  

\journal{Nuclear Physics A} % do not change
\begin{document} % do not change

\begin{frontmatter} % do not change

%% QM09Author: please enter your  
%% Title, author and address info here; please do not use footnotes

% Your Title - please modify
\title{Applicability of viscous hydrodynamics at RHIC}

% Principle author, and co-authors - please modify
\author{Denes Molnar$^{a,b}$ and Pasi Huovinen$^{a,c}$}

% Address - please modify
% note that if you have authors from several institutions, we recommend
% labelling these [a], [b], [c] etc.
\address[a]{Physics Department, Purdue University,
525 Northwestern Avenue, West Lafayette, IN 47906, USA}
\address[b]{RIKEN BNL Research Center, Brookhaven National Laboratory,
Upton, NY 11973, USA}
\address[c]{Institute f\"ur Theoretische Physik, 
Johann Wolfgang Goethe-Universit\"at, Frankfurt am Main, Germany}

\begin{abstract} % do not change
%% Text of abstract goes here - please modify
In an earlier work\cite{isvstr0} we established that causal Israel-Stewart 
viscous hydrodynamics
is only accurate in RHIC applications at very low 
shear viscosities $4\pi \eta_s/s \lton 1.5-2$.
We show here that the region of applicability is 
 significantly reduced if bulk viscosity plays a role in the dynamics.
\end{abstract} % do not change

\end{frontmatter} % do not change

%% QM09: we keep linenumbers at least for initial version
%\linenumbers % do not change

%% start of main text - please modify. Below is a sub-set (single section) 
%% of an earlier proceedings that show how one can handle references 
%% and figures etc.
%%\section{}\label{}

\section{Introduction}

There has been a lot of recent interest in quantifying the effect of 
viscosity on observables in heavy-ion 
collisions at the Relativistic Heavy Ion Collider (RHIC) at Brookhaven.
Calculations are most commonly performed using causal dissipative 
hydrodynamics\cite{romatschkev2,songv2,teaneyv2,ISv2},
though covariant transport theory can also be
utilized\cite{ISv2,minvisc,Frankfurtvisc}.

Hydrodynamics assumes that the system is
near local thermal equilibrium. Its region of applicability 
can only be reliably 
determined with the help of a fully nonequilibrium theory.
In a recent work\cite{isvstr0}
we used covariant transport to 
establish the region of validity for the causal 
Israel-Stewart (IS) formulation of viscous hydrodynamics, in a longitudinally
boost invariant setting with a massless $e = 3p$ equation
of state (EOS) and only shear viscosity. We found that for typical conditions 
expected in nuclear collisions at RHIC, IS
hydrodynamics is a very good
approximation (more precisely, 10\% accurate in computing dissipative effects)
when the shear viscosity ($\eta_s$) to entropy density ($s$) ratio is not too 
large, $4\pi \eta_s/s \lton 1.5-2$. A useful rule of thumb we obtained is
that, in order to reach such accuracy,
dissipative corrections to pressure and entropy must not exceed about 20\%.
This is only a necessary condition but its main advantage is that it
can be tested directly from the hydrodynamic calculation.

Here we study the region of validity of IS hydrodynamics 
for systems with shear {\em and} bulk viscosity.
Quantifying bulk effects in covariant transport near the
hydrodynamic limit is unfortunately unfeasible,
at least for a one-component system with $2\to 2$ scattering, because
bulk viscosity is at least two orders of magnitude smaller 
than shear viscosity. In order to proceed, we assume that 
the $20$\% rule of thumb above applies in the more general ``shear+bulk'' 
case as well.

Because for a massless equation of state bulk viscosity identically vanishes,
we here use a more realistic result from lattice QCD\cite{latticeEOS2007}
(we smoothly merge the EOS onto that of a hadron gas at low 
$T \lton 160$ MeV). Reliable calculations of viscosity in QCD are 
unfortunately not available for temperatures relevant for RHIC. We therefore
focus on the ``minimal viscosity''\cite{SonLimit} 
paradigm, i.e., set $\eta_s = s/(4\pi)$. 
For the bulk viscosity, we consider a Lorentzian in temperature
$\zeta = \zeta_m\, s /[1+(T-T_c)^2/\Delta T^2]$, where the peak height and 
width
are adjustable parameters. Matching to 
the bulk viscosity estimate from Meyer\cite{Meyerzetas}
gives our {\em default parameterization}
$\zeta_m = 0.2$, $\Delta T = 0.03$ GeV and $T_c = 0.192$ GeV.
Though our approach 
is similar to an earlier study by Fries et al\cite{FriesBulk}, 
{\em the main difference
is that we also ensure thermodynamic consistency because we use the 
complete set of
Israel-Stewart equations of motion}. We also map out a much wider range of
initial conditions.
\begin{figure}[ht]
\centering
\epsfysize=5cm
\epsfbox{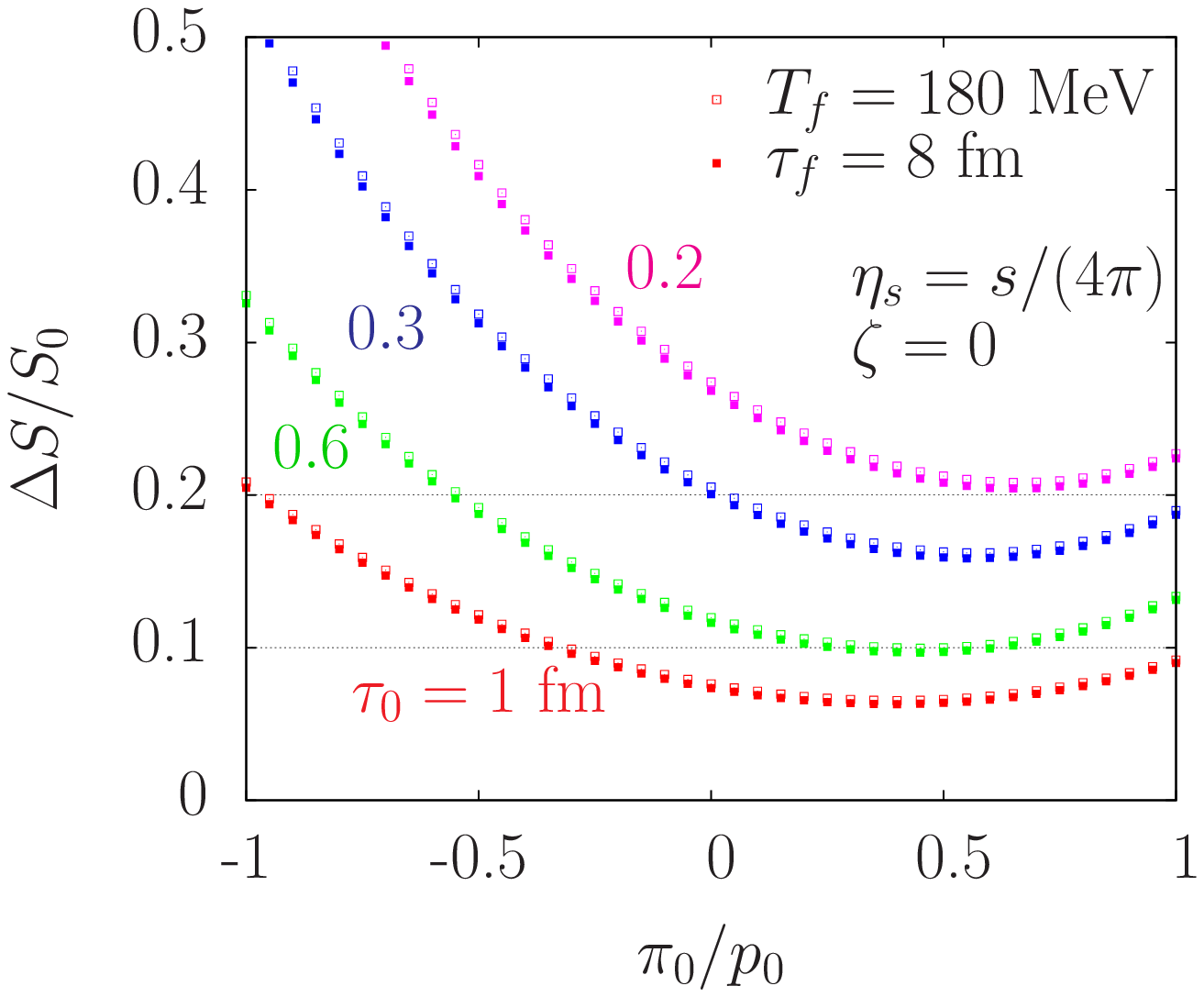}
\epsfysize=5cm
\epsfbox{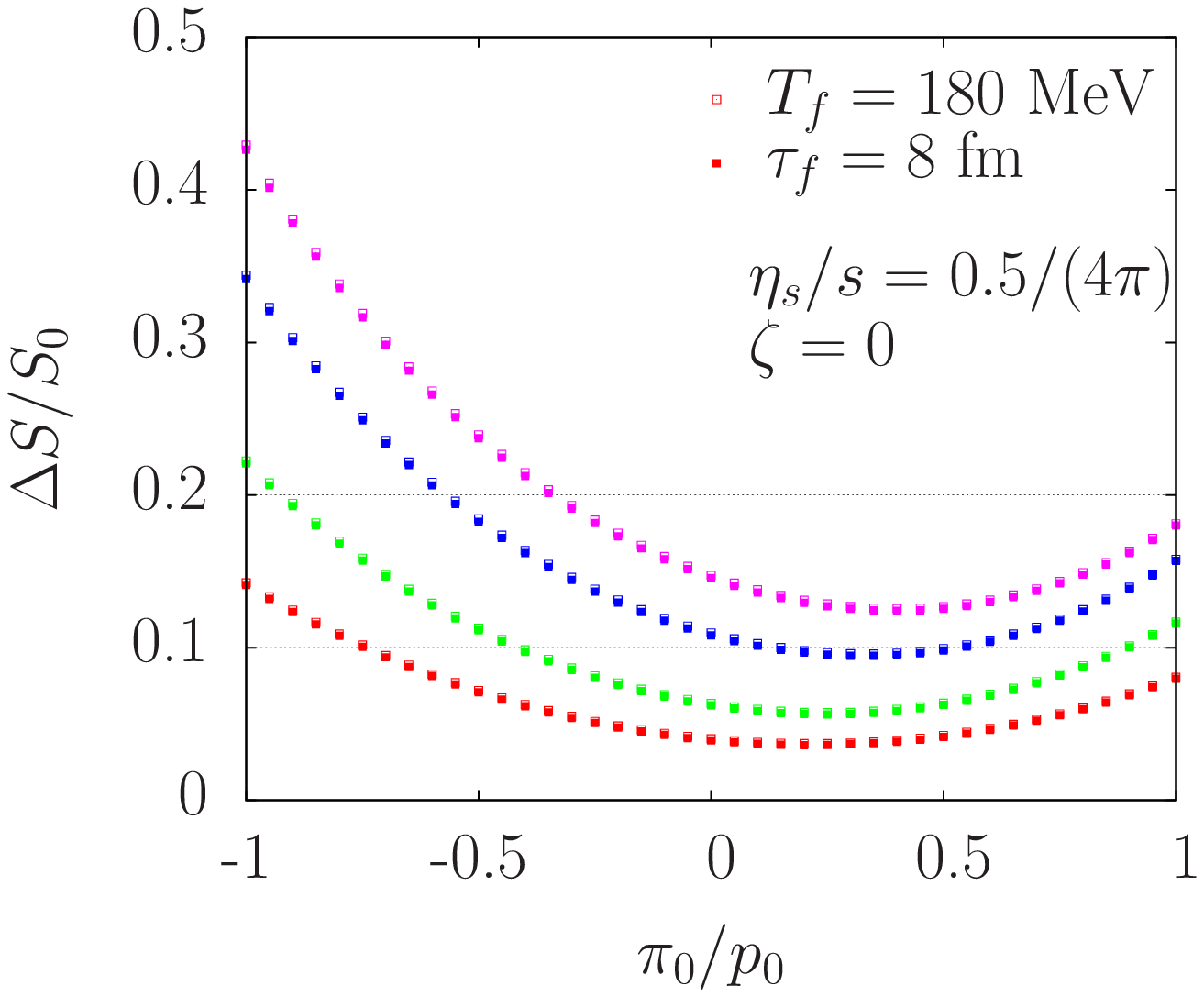}
\caption[]
{Entropy produced by the time the system cools down to $T = 180$ MeV 
($\tau \approx 8$ fm)
relative to the initial entropy, plotted as a function of 
thermalization time and initial shear stress, for $\eta_s/s = 1/(4\pi)$ (left)
and $0.5/(4\pi)$ (right).}
\label{Fig:1}
\end{figure}

\section{Main results}

We solve the complete set of Israel-Stewart equations (cf. \cite{isvstr0})
for a boost-invariant Bjorken scenario with axial and transverse translational
symmetry. We only highlight here the equation 
for bulk pressure
\be
\tau_\Pi \frac{d \Pi}{d\tau} = 
    -\left(\Pi + \frac{\zeta}{\tau}\right) 
    - \frac{\Pi\, \tau_\Pi}{2} 
        \left(\frac{1}{\tau} 
              + \frac{d}{d\tau} \ln \frac{\zeta}{\tau_\Pi\, T}\right) \ .
\ee
Here $\tau_\Pi$ is bulk pressure relaxation time.
If the last $\Pi\tau_\Pi...$ term is ignored, the entropy
generation rate per unit rapidity 
$d(dS/d\eta)/d\tau = \tau A_T [\Pi^2/(\zeta T) 
+ 3\pi_L^2/(4\eta_s T)]$ 
becomes inconsistent with the Israel-Stewart expression for entropy
$dS/d\eta = \tau A_T [s_{eq} - \Pi^2 \tau_\Pi/(2T\zeta) 
                             - 3\pi_L^2 \tau_\pi/(4T\eta_s)]$
($\pi_L$ is the shear correction to the longitudinal pressure, 
$\tau_\pi$ is the shear stress relaxation time, $A_T$ is the 
transverse area of the system, and $s_{eq}$ is the entropy density in local
equilibrium).
{\em Therefore, entropy production was overestimated in \cite{FriesBulk}.}

Motivated by kinetic theory we set
$\tau_\pi = 6\eta_s/(sT)$, but for simplicity 
take $\tau_\Pi = \tau_\pi$ (in kinetic theory 
$\tau_\Pi \approx 5\tau_\pi/3$ near the massless limit).
Our default initial condition for Au+Au at RHIC 
is $e_0 = 15$~GeV/fm$^3$ ($T_0 \approx 0.297$~GeV)
at a thermalization time $\tau_0 = 0.6$~fm. For $\tau_0 = 0.3$ and $1$~fm, 
we scale on an
isentropic curve $\tau_0 s_0 = const$. For initial shear stress and 
bulk pressure we map out wide ranges $-p_0<\pi_0 <p_0$,
$-p_0/2 < \Pi_0 < p_0/2$ where $p_0$ is the initial pressure.
Three choices are of special interest: i) local equilibrium (LTE)
$\pi_0 = \Pi_0 = 0$;
ii) Navier-Stokes (NS) 
$\pi_0 = -(4/3)\eta_s(T_0)/\tau_0$, $\Pi_0 = - \zeta(T_0)/\tau_0$;
and iii) gluon saturation\cite{CGC} (CGC)
$\pi_0 \approx -p_0$, $\Pi_0 \approx 0$.

The main quantity we analyze is the entropy produced relative to the initial
entropy $\Delta S / S_0$. We shall impose $\Delta S/S_0 \lton 0.2$ as
the condition for region of validity. 
Because we only consider longitudinal expansion, 
we focus on entropy production until the beginning of the hadronic
phase $T_f = 180$~MeV
(this is almost identical to entropy production until $\tau_f = 8$ fm).
Figure~\ref{Fig:1}(left) 
shows our results for ``minimal'' shear and vanishing bulk 
viscosity. For LTE initial conditions, IS hydrodynamics is 
applicable when $\tau_0 \gton 0.3$~fm. Good accuracy for NS initial conditions
requires $\tau_0 \gton 0.6$~fm ($\pi_0^{NS}/p_0 
\approx -0.846$, $-0.585$, and $-0.488$ for $\tau_0 = 0.3$, $0.6$, 
and $1$~fm), 
while for CGC initial conditions even later $\tau_0 \gton 1$~fm.

\begin{figure}[ht]
\centering
\epsfysize=4.5cm
\epsfbox{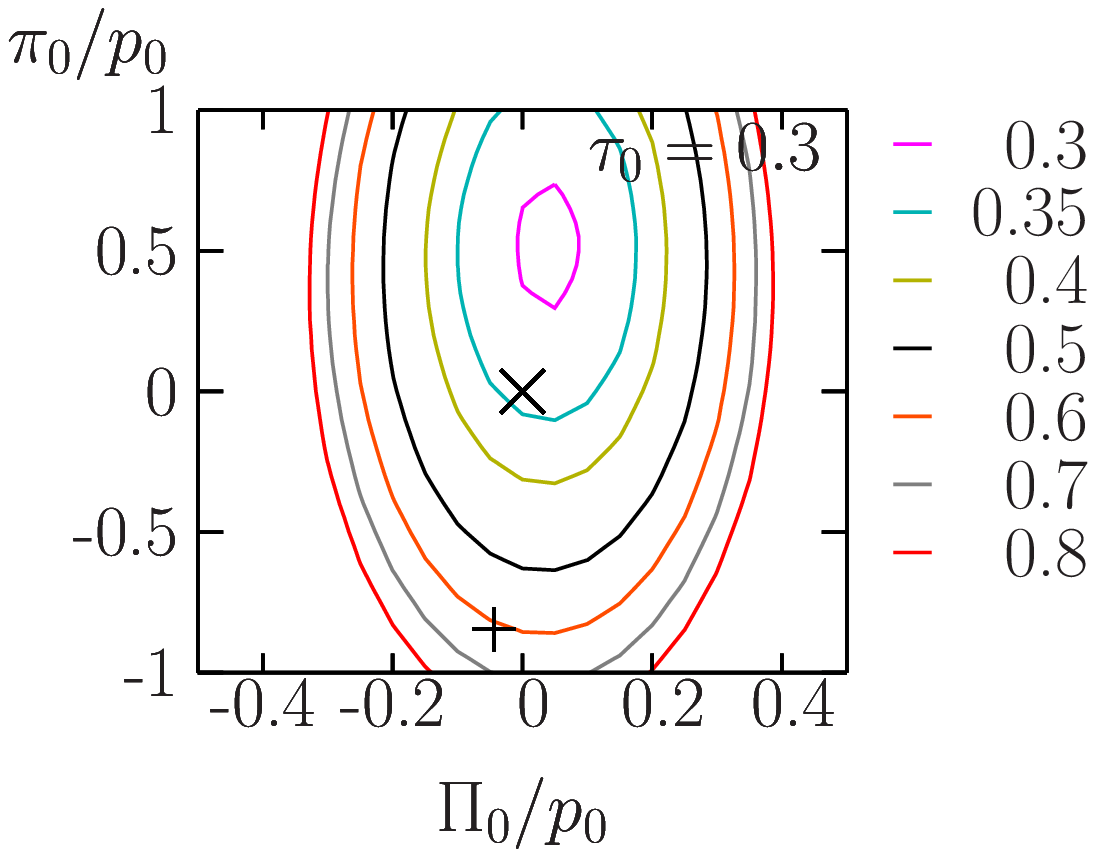}
\epsfysize=4.5cm
\epsfbox{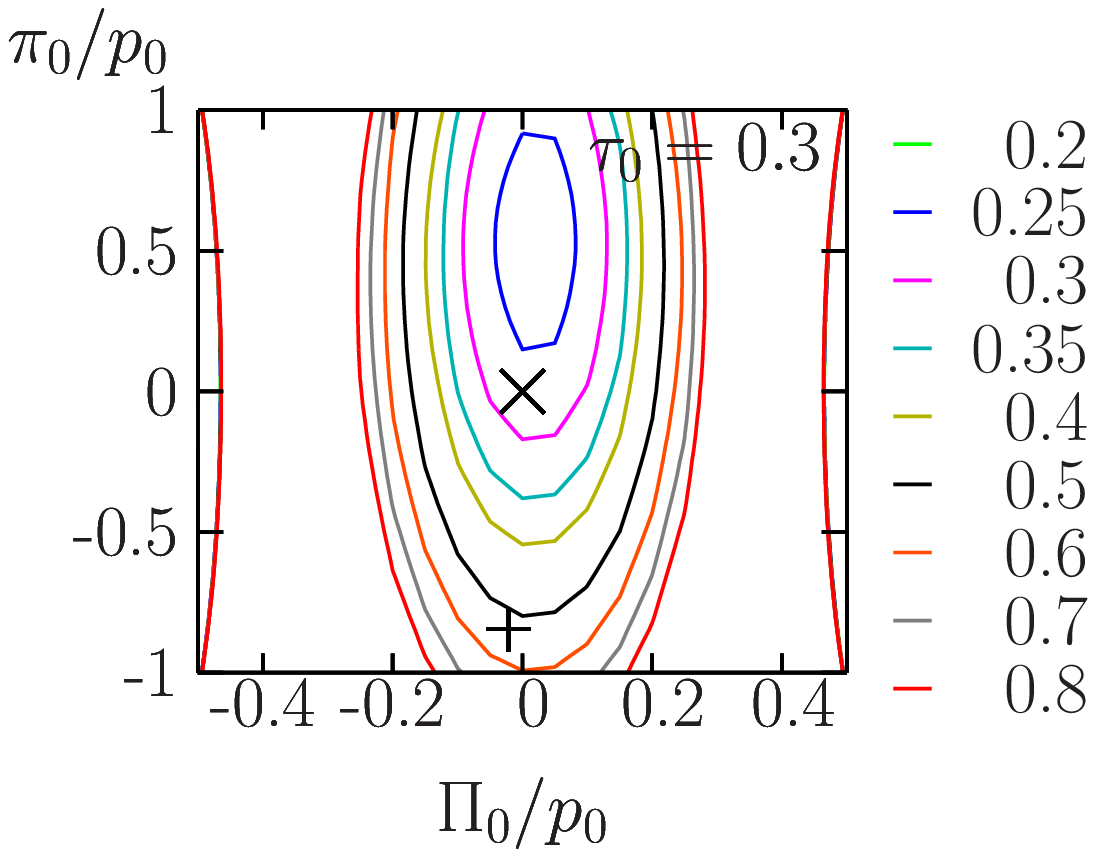}\\
\centering
\epsfysize=4.5cm
\epsfbox{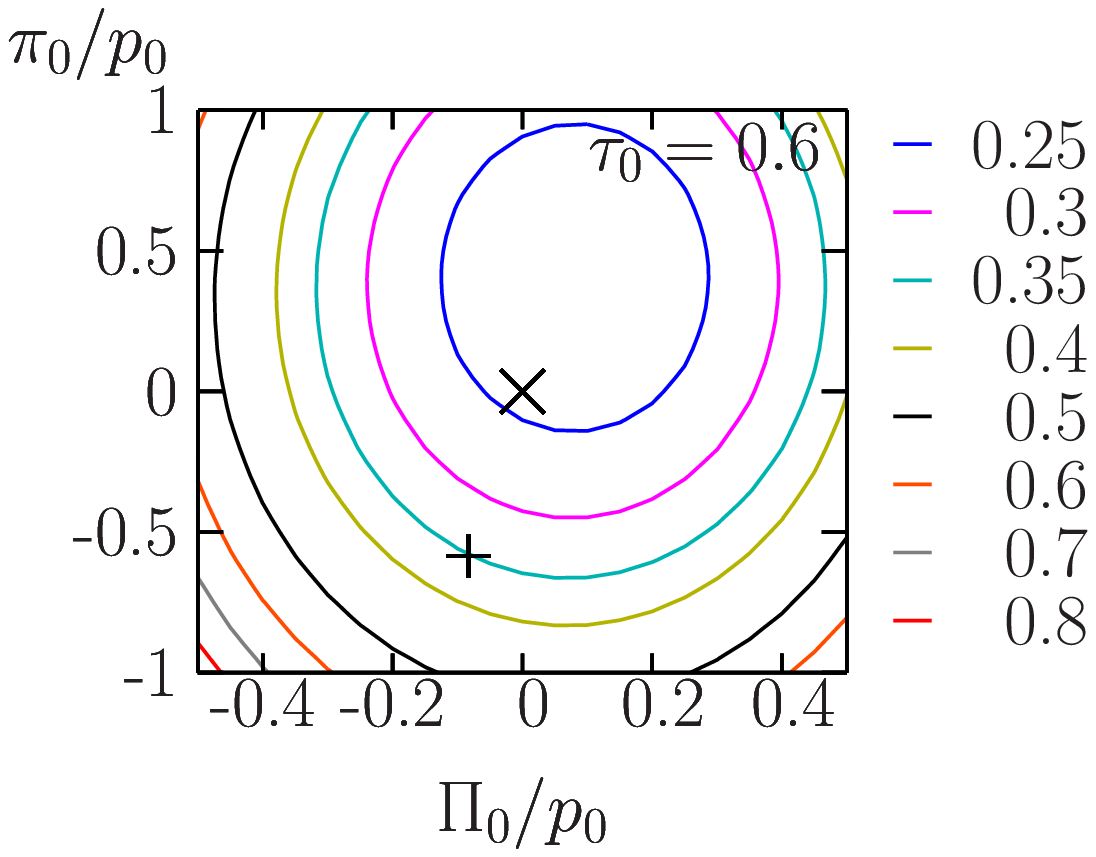}
\hspace*{-0.8cm}
\epsfysize=4.5cm
\epsfbox{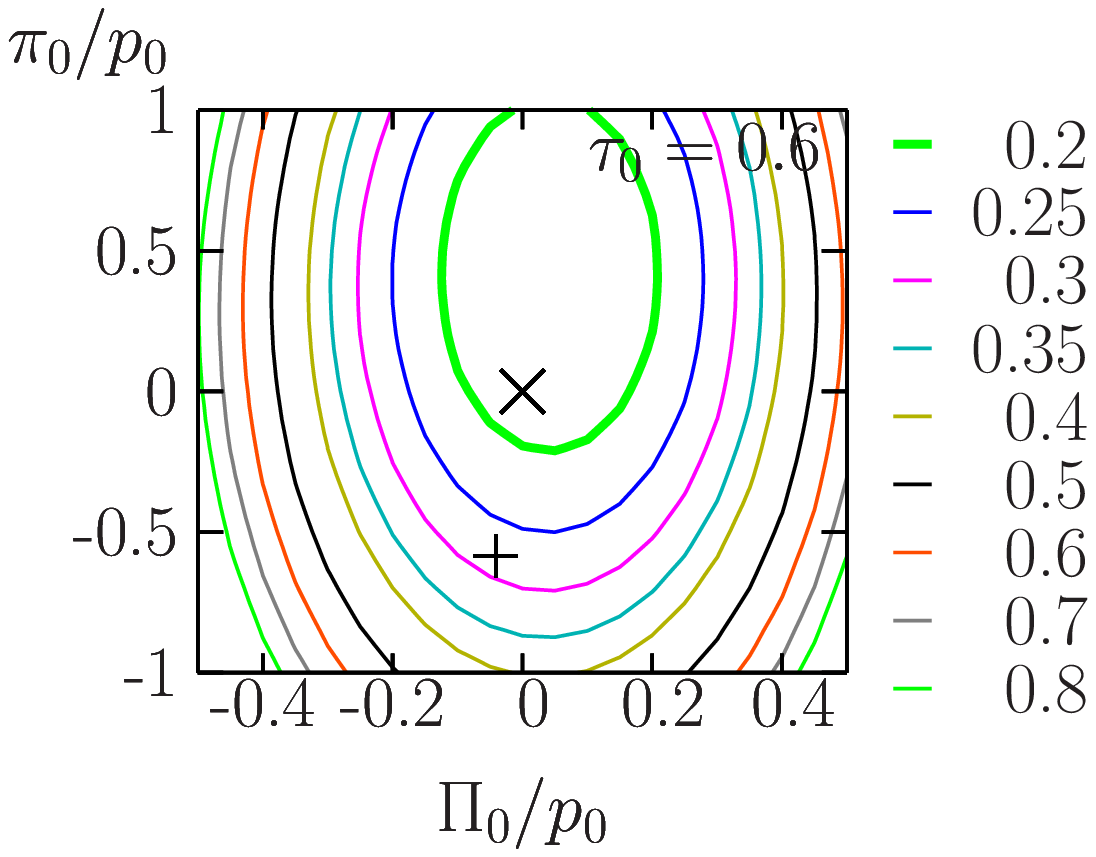}\\
\centering
\epsfysize=4.5cm
\epsfbox{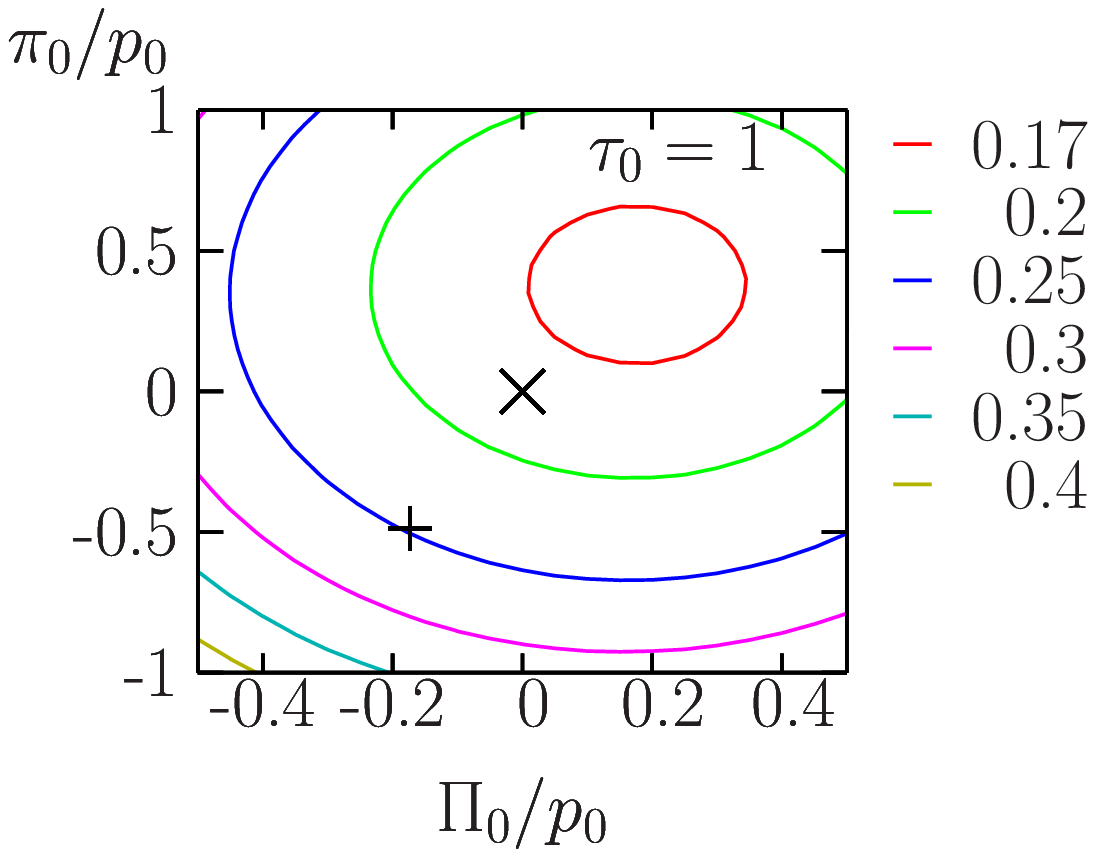}
\epsfysize=4.5cm
\epsfbox{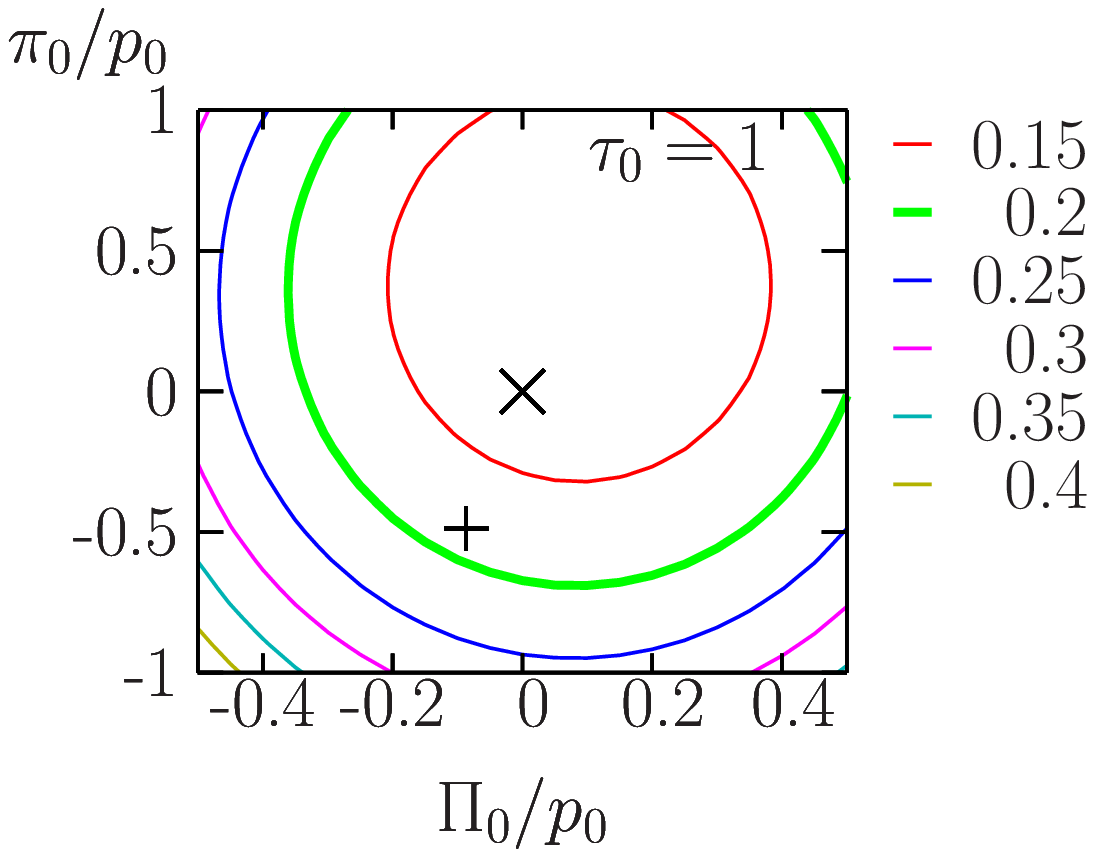}\\
\caption[]{Relative entropy production until $T_f = 180$~MeV
as a function of initial shear stress and bulk pressure, for thermalization
times $\tau_0 = 0.3$ (top), $0.6$ (middle), and $1$~fm (bottom).
Left column is for $\zeta(T)$ based on Meyer's calculation\cite{Meyerzetas},
right column is for half that large $\zeta(T)$. 
Crosses indicate local equilibrium initial conditions, while pluses
are for Navier-Stokes.
In all cases $\eta_s/s = 1/(4\pi)$.}
\label{Fig:2}
\end{figure}

\begin{figure}[ht]
\centering
\epsfysize=5cm
\epsfbox{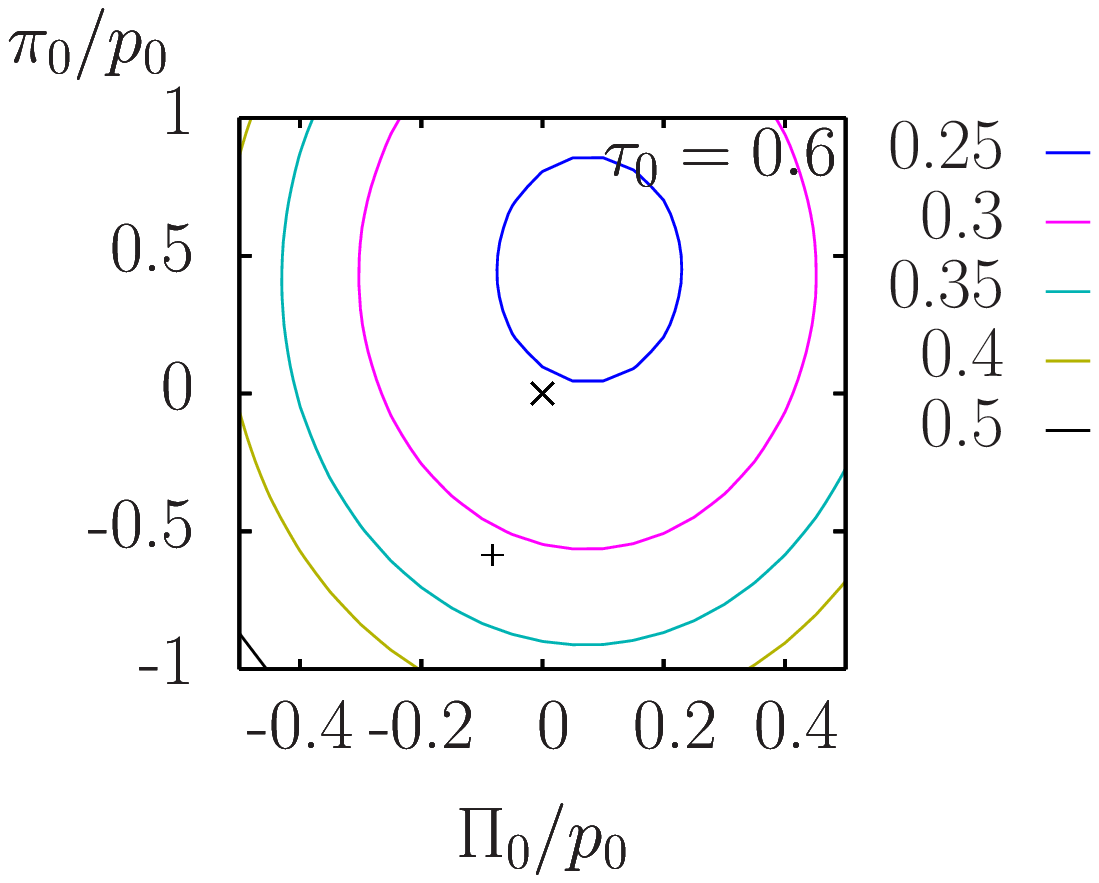}
\epsfysize=5cm
\epsfbox{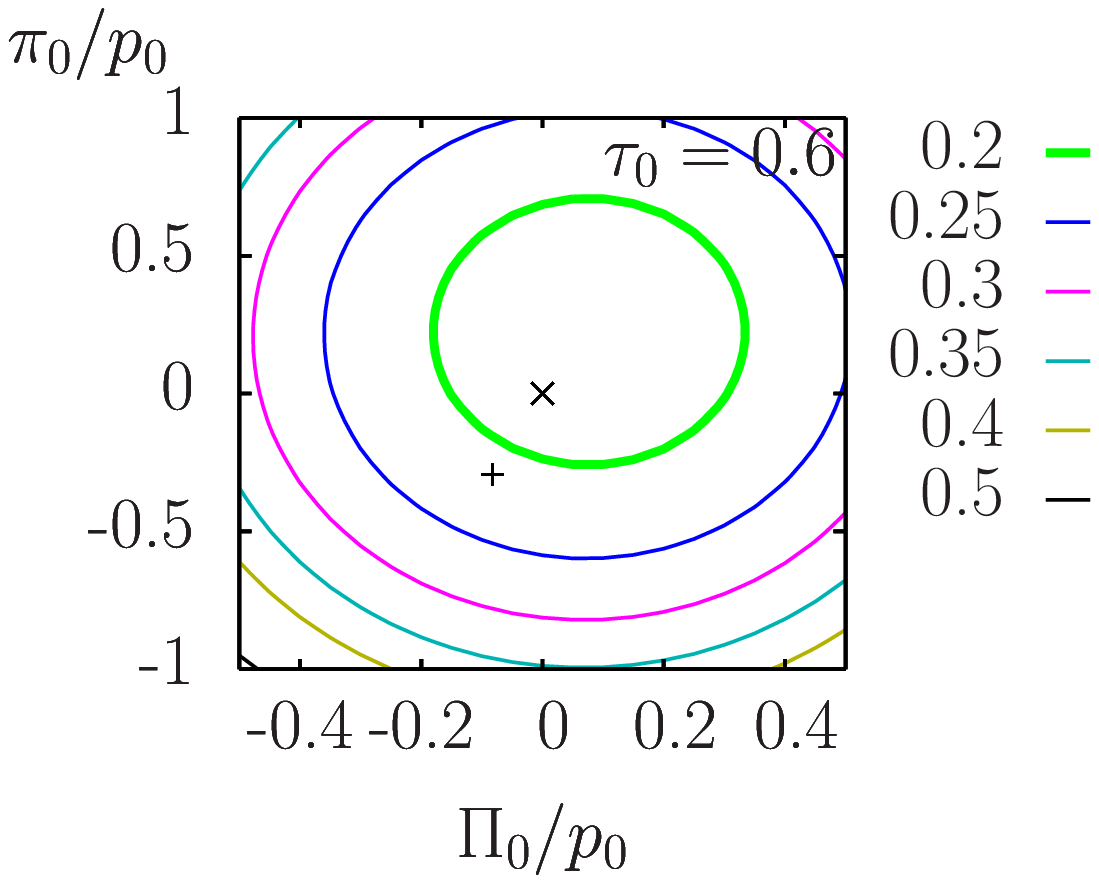}
\caption[]{Same as Fig.~\ref{Fig:2} with our default parameters
but halved relaxation times (left), or $\eta_s/s=0.5/(4\pi)$ (right).}
\label{Fig:3}
\end{figure}

Let us now turn on bulk viscosity. The left column of Figure~\ref{Fig:2}
shows $\Delta S/S_0$ for our default $\zeta(T)$ parameterization.
Due to the additional entropy produced, 
LTE initial conditions now require $\tau_0 \gton 1$~fm, while
simulations from NS and CGC ones necessitate even later thermalization.
The situation improves somewhat if bulk viscosity is half as large
as our default ($\zeta_m = 0.1$). As seen in the right column,
LTE initial conditions are then suitable when $\tau_0 \gton 0.6$~fm, while
NS can be accommodated if $\tau_0 \gton 1$~fm. Results are quite similar
if instead the width of the $\zeta(T)$ peak is halved, i.e., $\zeta_m = 0.2$, 
$\Delta T = 0.015$~GeV. 
We checked that shorter relaxation times 
help only a little,
as illustrated in Figure~\ref{Fig:3}(left) for $\tau_0 = 0.6$~fm.
On the other hand, the region of applicability widens substantially
if the shear viscosity is a factor of two smaller
(cf. Fig.~\ref{Fig:3}(right)).

\section{Conclusions}

We study the region of validity of Israel-Stewart viscous hydrodynamics
for conditions expected at RHIC, based on the entropy produced during the 
evolution when both shear and bulk viscosity are present.
Our results indicate that, unlike it was previously hoped, 
viscous hydrodynamics does not extend the range of dynamical description
to proper times earlier than $\tau \approx 0.6$~fm, even for local
thermal equilibrium initial conditions - {\em unless}
the influence of bulk viscosity is basically negligible, or 
the shear viscosity of hot and dense 
quark-gluon matter is significantly below the conjectured ``minimal'' 
value of $\eta_s = s/(4\pi)$.

%% end of main text

\section*{Acknowledgments} % please check/modify, comment out or delete if 
We thank RIKEN, 
Brookhaven National Laboratory and
the US Department of Energy [DE-AC02-98CH10886] for providing facilities
essential for the completion of this work.

 % do not change 
\end{document}